\documentclass{article}

\usepackage{arxiv}
\usepackage{wrapfig}
\usepackage{tabularx, booktabs}
\usepackage{graphicx}
\usepackage{multirow}
\usepackage{setspace}
\usepackage{xcolor}
\usepackage{subcaption}
\usepackage{array}
\usepackage{balance}
\usepackage{enumitem}
\usepackage{balance}
\usepackage{caption}
\usepackage{bm}
\usepackage{siunitx}
\usepackage{amsfonts}       % blackboard math symbols
\usepackage[linesnumbered,ruled,vlined]{algorithm2e}

\SetKwInput{KwTarget}{Target}

\newcolumntype{C}[1]{>{\centering\arraybackslash}p{#1}}
\newcolumntype{R}[1]{>{\raggedleft\arraybackslash}p{#1}}
\newcolumntype{L}[1]{>{\raggedright\arraybackslash}p{#1}}

\makeatletter
\newcommand{\ssymbol}[1]{^{\@fnsymbol{#1}}}
\makeatother

\title{Learning To Retrieve: How to Train a Dense Retrieval Model Effectively and Efficiently}

\author{
    Jingtao Zhan \\
    BNRist, DCST, Tsinghua University \\
    Beijing, China \\
    \texttt{jingtaozhan@gmail.com} 
    \And
    Jiaxin Mao \\
    BNRist, DCST, Tsinghua University \\
    Beijing, China \\
    \texttt{maojiaxin@gmail.com} 
    \And
    Yiqun Liu\thanks{Corresponding author} \\
    BNRist, DCST, Tsinghua University \\
    Beijing, China \\
    \texttt{yiqunliu@tsinghua.edu.cn} \\ 
    \And
    Min Zhang \\
    BNRist, DCST, Tsinghua University \\
    Beijing, China \\
    \texttt{z-m@tsinghua.edu.cn} \\
    \And
    Shaoping Ma \\
    BNRist, DCST, Tsinghua University \\
    Beijing, China \\
    \texttt{msp@tsinghua.edu.cn} \\  
}

\begin{document}
\maketitle

\begin{abstract}
Ranking has always been one of the top concerns in information retrieval research. For decades, lexical matching signal has dominated the ad-hoc retrieval process, but it also has inherent defects, such as the vocabulary mismatch problem. Recently, Dense Retrieval (DR) technique has been proposed to alleviate these limitations by capturing the deep semantic relationship between queries and documents. The training of most existing Dense Retrieval models relies on sampling negative instances from the corpus to optimize a pairwise loss function. Through investigation, we find that this kind of training strategy is biased and fails to optimize full retrieval performance effectively and efficiently. To solve this problem, we propose a Learning To Retrieve (LTRe) training technique. LTRe constructs the document index beforehand. At each training iteration, it performs full retrieval without negative sampling and then updates the query representation model parameters. Through this process, it teaches the DR model how to retrieve relevant documents from the entire corpus instead of how to rerank a potentially biased sample of documents. Experiments in both passage retrieval and document retrieval tasks show that: 1) in terms of effectiveness, LTRe significantly outperforms all competitive sparse and dense baselines. It even gains better performance than the BM25-BERT cascade system under reasonable latency constraints. 2) in terms of training efficiency, compared with the previous state-of-the-art DR method, LTRe provides more than 170x speed-up in the training process. Training with a compressed index further saves computing resources with minor performance loss.
\end{abstract}

\section{introduction}

Retrieving relevant documents is essential in many tasks, such as question answering~\cite{guu2020realm}, web search~\cite{craswell2020overview, bajaj2016ms}. Currently, most search systems\cite{nogueira2019passage, nogueira2019multi, yan2019idst, dai2019deeper} adopt a pipeline method: an efficient first-stage \emph{retriever} retrieves a small set of documents from the entire corpus, and then a powerful but slow second-stage \emph{ranker} reranks them.

 Traditionally, information retrieval utilizes lexical retrieval models for first-stage retrieval, such as BM25~\cite{robertson1994some}. Though still widely adopted today, they rely on exact matching and ignore semantic meanings or other high-level properties. For example, they may fail if the query and document use different terms for the same meaning, known as the \emph{vocabulary mismatch problem}. However, the accuracy of first-stage retrieval is vital for searching systems. Several works ~\cite{han2020learning, nogueira2019document} confirm that a better first-stage retriever significantly improves ranking performance.

A promising alternative approach is Dense Retrieval (DR)~\cite{zhan2020repbert, luan2020sparsedense, lee2019latent, chang2020pre, gao2020complementing, karpukhin2020dense}, which matches queries and documents in a low-dimension embedding space and is as fast as traditional methods~\cite{xiong2020approximate, zhan2020repbert}. It usually utilizes deep neural networks to learn the low-dimensional representations of queries and documents, which contain rich information, such as semantic meanings, sentence structures, language styles, etc. Given a query representation and a document representation, the inner product or cosine similarity is regarded as the relevance score. 

How to effectively and efficiently train a DR model is still an open question. Previous works utilize the negative sampling method. Given a query, they sample several irrelevant documents from the corpus and mix them with relevant ones. Then they train the DR model to rerank these documents. Different sampling methods are explored, such as BM25 top documents~\cite{gao2020complementing}, in-batch negatives~\cite{zhan2020repbert, karpukhin2020dense}, random samples~\cite{huang2020embedding}, asynchronous approximate nearest neighbors (Async ANN)~\cite{xiong2020approximate, guu2020realm}, and combinations of several above techniques~\cite{luan2020sparsedense}.

However, there exists discrepancy between training and inference when using negative sampling methods. Such methods teach a model to rerank the sampled documents during training, but require the model to retrieve from the entire corpus during inference. In other words, training a model to \emph{rerank} is not equivalent to training it to \emph{retrieve}. For example, the model may hardly learn to retrieve if the negatives are too weak; it may be optimized in the wrong direction if the negative samples are biased. In both situations, the model can successfully rerank the sampled documents but cannot effectively retrieve the most relevant documents from the whole corpus.

Our experiments confirm the aforementioned discrepancy. To resolve it, we present \textbf{L}earning \textbf{T}o \textbf{Re}trieve (\textbf{LTRe}), an effective and efficient training mechanism for Dense Retrieval. 
LTRe first uses a pretrained document encoder to represent documents as embeddings, which are fixed throughout the training process. At each training step, the DR model outputs a batch of queries’ representations. LTRe then uses them and performs full retrieval. Based on the retrieval results, LTRe updates the model parameters. 
Compared with previous works, LTRe teaches the DR model to retrieve rather than to rerank. Thus it is consistent in training and inference. We investigate the training procedure and verify that LTRe leads to fast and steady optimization.

\begin{figure}
	\centering
    \includegraphics[width=0.35\linewidth, keepaspectratio=True]{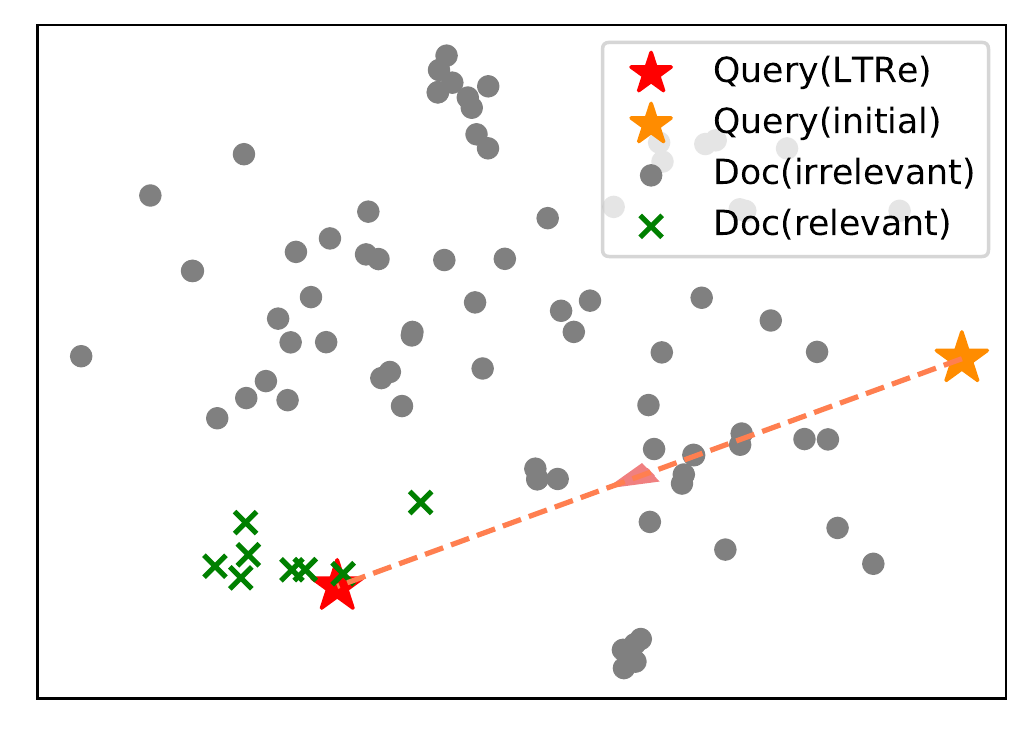}
    \caption{The t-SNE plot of query and document representations. Query(initial) and Query(LTRe) correspond to query representations before and after training, respectively. Qid: 1129237; Query: `hydrogen is a liquid below what temperature' from TREC 2019 DL Track document retrieval test set.
    } 
    \label{fig:tsne}
\end{figure}

From a higher perspective, LTRe teaches the DR model to represent the user's information need. 
The DR model maps both corpus and queries to the same representation space. The goal is to learn a mapping function that maps queries close to the relevant documents and far from the irrelevant ones.
To achieve this, LTRe runs as follows. 
It adopts a pretrained document encoder to map the corpus before training. 
According to the entire corpus distribution in the representation space, LTRe optimizes the query mapping globally.
We visualize an example with t-SNE~\cite{maaten2008visualizing} in Figure~\ref{fig:tsne}. After training, the DR model better understands the user's information need and maps the query closer to the relevant documents. Thus, the retrieval performance is significantly improved.

We conduct experiments on TREC 2019 DL Track~\cite{craswell2020overview} passage retrieval and document retrieval tasks. The results show that:
\begin{itemize}
	\item \textbf{First-stage Effectiveness:} LTRe significantly outperforms competitive sparse and dense retrieval baselines. Under reasonable latency constraints, it outperforms the BM25-BERT cascade system.
	\item \textbf{Two-stage Effectiveness:} Using LTRe as the first-stage retriever significantly improves the second-stage reranking performance. LTRe+BERT cascade system outperforms a recent proposed competitive end-to-end retrieval method~\cite{Khattab2020ColBERTEA} in both effectiveness and efficiency.
	\item \textbf{Training Efficiency:} Compared with the previous state-of-the-art DR training method~\cite{xiong2020approximate}, LTRe provides more than 170x speedup in training time. Training with a compressed index substantially saves computing resources (4 gpus to 1 gpu) with minor performance loss.
\end{itemize} 

The remainder of this paper is organized as follows. 
We review related work in Section~\ref{sec:related_work} and present the DR background in Section~\ref{sec:overview}.
Then we describe our proposed Learning To Retrieve method in Section~\ref{sec:ltre}.
We show our experimental setup and results in Sections~\ref{sec:exp_setting} and \ref{sec:results_discuss}.
Finally, we conclude this work and suggest future directions in Section~\ref{sec:conclusion}.

\section{Related Work}
\label{sec:related_work}

Recent studies use neural networks to improve first-stage retrieval.
We classify them into three categories, namely sparse retrieval, dense retrieval, and end-to-end retrieval.

\textbf{Sparse Retrieval:} 
Several works use neural networks to improve sparse retrieval performance.
doc2query~\cite{nogueira2019document} and docTTTTTquery~\cite{nogueira2019doc2query} are proposed to alleviate the \emph{vocabulary mismatch problem} where the query terms are different from those in the relevant documents. They use deep language models~\cite{devlin2019bert, raffel2020exploring} to predict possible query terms for each document. They expand the documents with these predicted query terms. 
DeepCT~\cite{dai2019context} replaced the term frequency field in BM25 with term weights predicted by BERT~\cite{devlin2019bert}. Thus, the bag-of-words retrievers use term weights based on semantic importance rather than term frequency.  

\textbf{Dense Retrieval:} Dense Retrieval is a representation-based first-stage retriever. It relies on neural models to represent text as embeddings, i.e., real-valued vectors in low-dimensional space. Similarity search~\cite{johnson2019billion}, such as maximum inner product search, is then used to efficiently retrieve the vectors that are similar to the query vector. Section~\ref{sec:overview} introduces its architecture and inference procedure in details.
Though early research~\cite{guo2019deep} found that representation-based models usually underperform interaction-based models, recent language modeling advances prompt researchers to revisit the Dense Retrieval approach. With pretrained deep language models, several works~\cite{zhan2020repbert, lee2019latent, chang2020pre, karpukhin2020dense} demonstrated that DR can significantly outperform traditional methods. We present the details of these works in Section~\ref{sec:existing_training_methods}. 

\textbf{End-to-End Retrieval:} Several works~\cite{Khattab2020ColBERTEA, nie2020dc, MacAvaney2020EfficientDR} focused on improving the inference-time efficiency of the BERT ranker~\cite{nogueira2019passage}. Khattab et al.~\cite{Khattab2020ColBERTEA} further applied their proposed ColBERT for end-to-end retrieval. Its retrieval performance almost matches existing BERT-based models and outperforms many non-BERT baselines. For each document, they precompute contextualized word representations offline. During online stage, they utilize light-weight interactions between query and document terms to judge relevance. Due to its interaction-based architecture during online retrieval, its latency (0.5 s/query) is too high to use a more sophisticated second-stage ranker, such as the BERT ranker.

\section{Overview of Dense Retrieval}
\label{sec:overview}

\begin{figure*}
    % Maximum length
    \hspace*{\fill}%
    \subcaptionbox{Negative sampling training method. For a given training query, it selects document samples and trains the DR model to rerank them.
    \label{fig:negative_flowchart}}
    {\includegraphics[height=.55\linewidth]{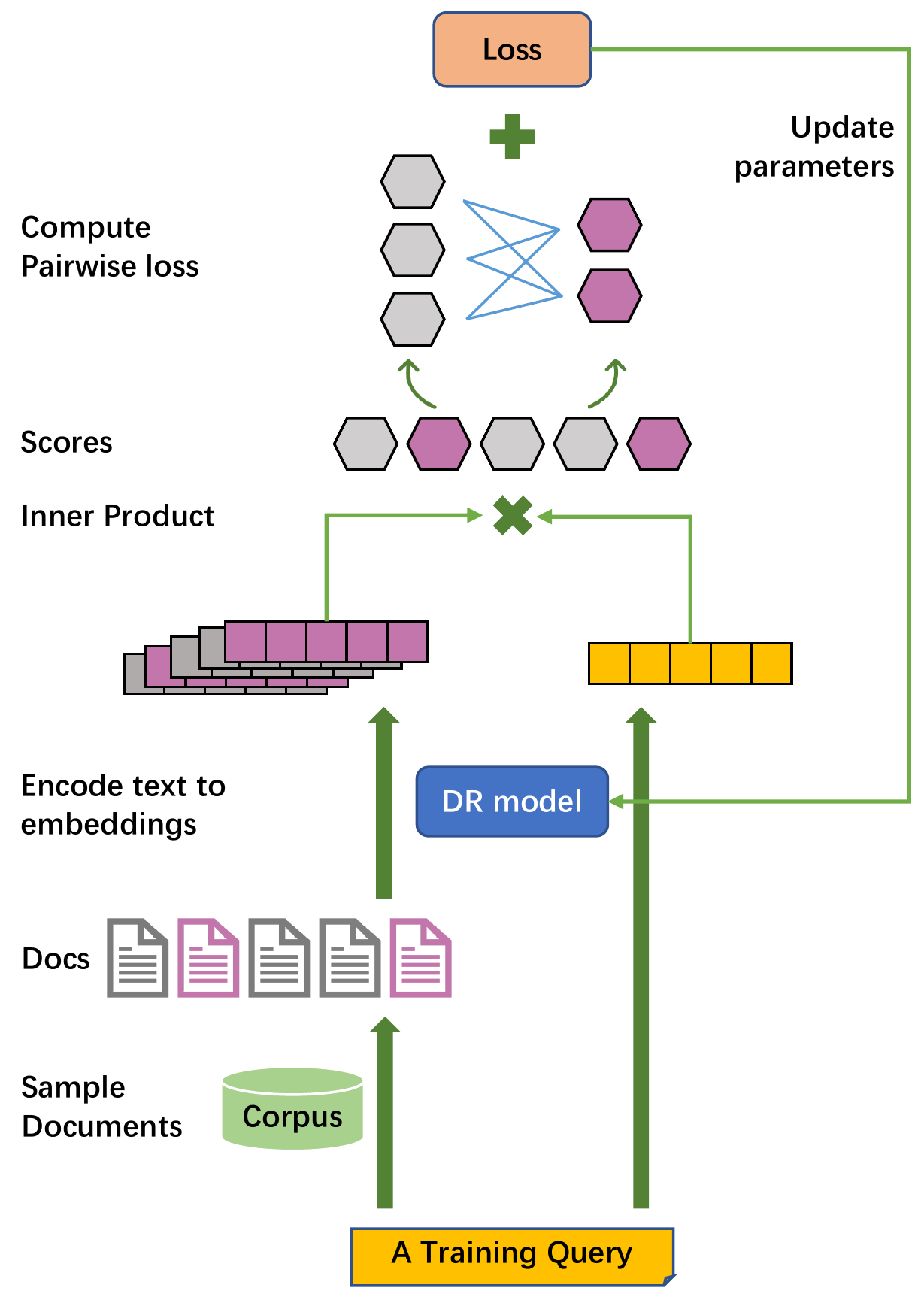}}
    \hfill%
    \subcaptionbox{Learning To Retrieve (LTRe). Each training iteration includes an inference procedure at the beginning. If the retrieved documents are all irrelevant, the last one is replaced with a relevant document.
    \label{fig:ltre_flowchart}}
    {\includegraphics[height=.55\linewidth]{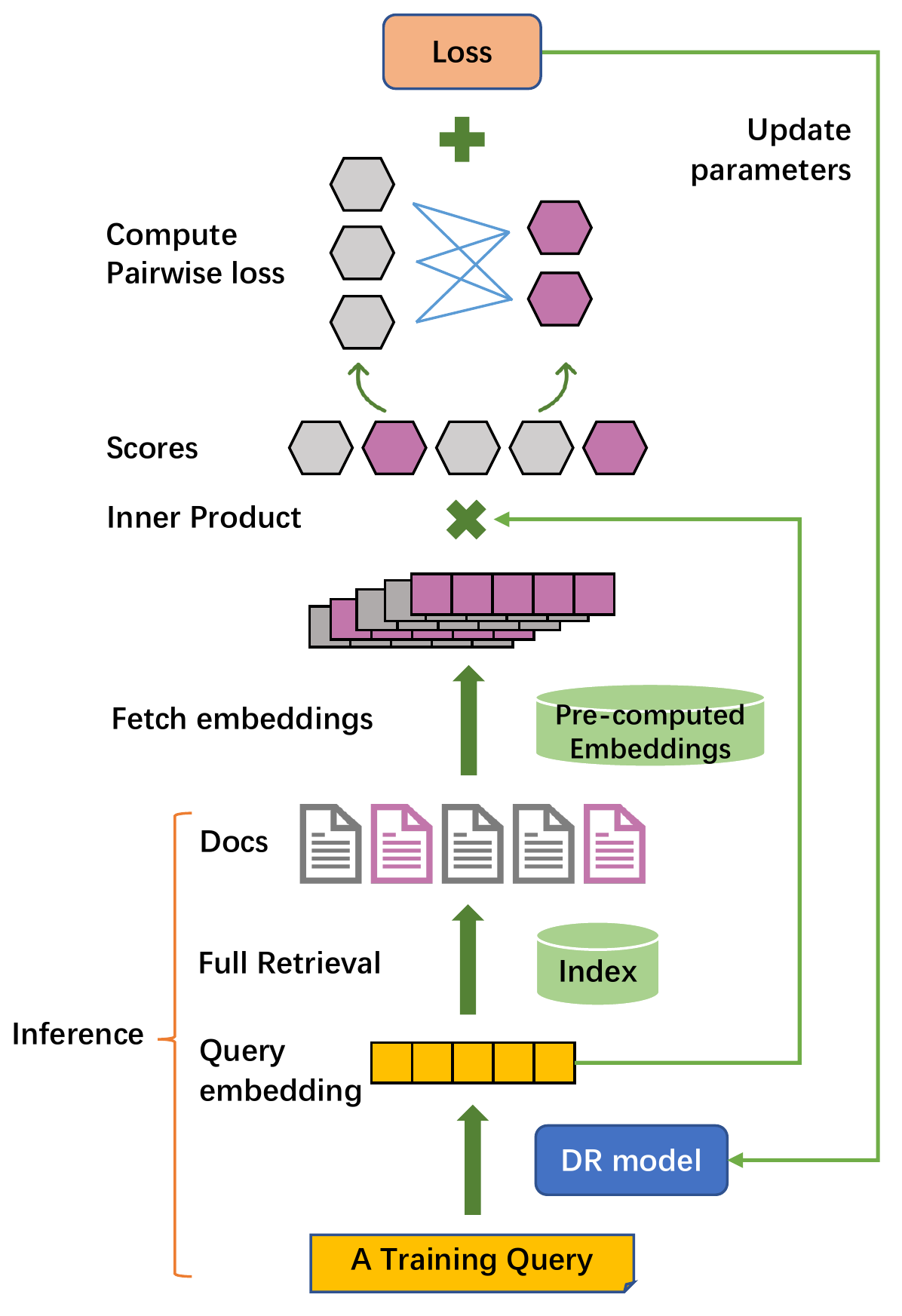}}%
    \hspace*{\fill}%
    \caption{The flow chart of negative sampling training method and our proposed Learning To Retrieve (LTRe).  Gray: document labeled irrelevant, Purple: document labeled relevant. The batch size is set to one.}
\end{figure*}

In this section, we introduce the background of Dense Retrieval. We present the architecture of DR model in Section~\ref{sec:DR_architecture} and show how DR model performs first-stage retrieval in Section~\ref{sec:DR_Inference}. Section~\ref{sec:existing_training_methods} introduces the existing training methods.

\subsection{Architecture}
\label{sec:DR_architecture}

\subsubsection{Text Encoder}
Dense Retrieval (DR) models use encoders to map documents and queries to k-dimensional real-valued vectors. Formally~\cite{chang2020pre}, given a query $q \in Q$ and a document $d \in D$, the DR model uses two functions, $ \varphi : Q \to \mathbb{R}^k $, $\psi : D \to \mathbb{R}^k $, to encode them to their associated representations $ \varphi(q) $ and $\psi(d)$, respectively. 

How to effectively represent text has been explored for years~\cite{hu2014convolutional, Huang2013LearningDS, Wan2016ADA}. Recent language modeling advances~\cite{devlin2019bert, beltagy2020longformer, liu2019roberta, dai2019transformer} provide many powerful representation tools. With these language models, several works~\cite{xiong2020approximate, karpukhin2020dense, guu2020realm} found that DR can significantly outperform traditional information retrieval methods . 

\subsubsection{Relevance Estimation}

Given a query representation $ \varphi(q) $ and a document representation $\psi(d)$, a relevance estimation function is formally defined as $f : \mathbb{R}^k \times \mathbb{R}^k \to \mathbb{R} $. Early studies~\cite{hu2014convolutional, Wan2016ADA} utilized multiple neural layers to learn an expressive function $f$. Such a time-consuming $f$ is applicable to the reranking task but impractical for full retrieval due to efficiency concerns. Recent studies~\cite{guu2020realm, xiong2020approximate, zhan2020repbert, gao2020complementing, luan2020sparsedense} employed inner product as a simple $f$ for the first-stage retrieval. This paper also follows this practice.

\subsection{Inference}
\label{sec:DR_Inference}

The DR model preprocesses the corpus offline. It represents the documents as embeddings and builds the index for fast search. During online inference, the DR model first encodes queries as embeddings and then retrieves documents from the entire corpus. Part of Figure~\ref{fig:ltre_flowchart} shows this inference procedure, which is annotated with a curly bracket.

Finding the nearest neighbors in the embedding space has been widely studied~\cite{ram2012maximum, johnson2019billion}. With a pre-built document index, the search is very efficient. Previous works~\cite{xiong2020approximate, zhan2020repbert} reported that DR is as efficient as traditional retrievers.

\subsection{Existing Training Methods}
\label{sec:existing_training_methods}
Previous works train the DR model to rerank the selected document samples. We classify them as the negative sampling training method and show the training process in Figure~\ref{fig:negative_flowchart}.
For a given training query, it selects several negative documents from the entire corpus. The negatives and the relevant documents form the document samples. The DR model encodes the queries and documents to embeddings and uses inner product to compute their relevance scores. Based on the annotations, the training method uses the scores to compute a pairwise loss. Through optimizing the pairwise loss, the DR model learns to rerank the document samples.

How to select negative samples is not straightforward, and different works use different strategies based on heuristics or empirical results. 
Huang et al.~\cite{huang2020embedding} used random negative samples because they believed it approximates the recall optimization task. 
Gao et al.~\cite{gao2020complementing} used the BM25 top documents as negatives. 
Karpukhin et al.~\cite{karpukhin2020dense} and Zhan et al.~\cite{zhan2020repbert} utilized a trick called in-batch negatives to reuse computation and reduce the computational cost. 
Xiong et al.~\cite{xiong2020approximate} provided a DR baseline using noise contrastive estimation (NCE)~\cite{gutmann2010noise}. It uses the highest scored negatives in batch.
Luan et al.~\cite{luan2020sparsedense} tried to combine several methods mentioned above. 

Another costly but well-performing training strategy, ANCE~\cite{xiong2020approximate}, used asynchronous approximate nearest neighbors (Async ANN) as negatives. Every thousands of training steps, it uses the current model parameters to re-encode and re-index the documents. Then it retrieves the top-n documents for the training queries, which are utilized as negatives for the following training iterations until the next refresh. In other words, the negatives are selected based on out-dated model parameters. Though it achieved a state-of-the-art first-stage retrieval performance, such training is very time-consuming. Section~\ref{sec:exp_train_efficiency_time} shows that it takes nearly a month till convergence with four gpus. 
While it is much more sophisticated than some simple methods, such as random sampling, we still classify Async ANN as a negative sampling method since it samples and fixes the negatives before each training segment begins.

\section{Learning To Retrieve}
\label{sec:ltre}

\subsection{Principles}
\label{sec:principles}

Before presenting our method, we propose two principles for an ideal DR training method. 

\begin{itemize}
	\item \textbf{Effectiveness:} The training strategy effectively improves the first-stage retrieval performance and benefits a second-stage ranker. 
	\item \textbf{Efficiency:} The training strategy is efficient and applicable to a large corpus. We consider both training time and computing resources~\footnote{Here we focus on the efficiency of training as the the efficiency in the inference-time is guaranteed by the maximum inner product search algorithms\cite{shen2015learning, johnson2019billion}.}.
\end{itemize} 

However, previous works at least break one rule, which will be discussed in our experiments. 

\subsection{Method}

Following the principles mentioned above, we present \textbf{L}earning \textbf{T}o \textbf{Re}trieve (\textbf{LTRe}), an effective and efficient Dense Retrieval training method. 

LTRe teaches the DR model to represent the user's information need. 
We show the training process in Figure~\ref{fig:ltre_flowchart}.
Before training, LTRe pre-computes the document embeddings with a pretrained document encoder and builds the index. They are fixed throughout the entire training process.
At each training iteration, it retrieves the top-n documents for a batch of queries, which is the same as inference time. 
Based on the retrieval results, LTRe optimizes the query representation process as follows.
It fetches the documents' pre-computed embeddings and computes the relevance scores. It then computes the pairwise loss and uses back-propagation to update the DR model parameters. The optimization of the pairwise loss forces the DR model to improve full retrieval performance.

The detailed process is shown in Algorithm~\ref{algo:LTRe}. Before training, it encodes documents to embeddings $E_{doc}$ and builds the index $I_{doc}$. During training, LTRe only updates the parameters of query encoder $\varphi$.

\begin{algorithm}[t]
\KwData{document embeddings $E_{doc}$, document index $I_{doc}$, training queries, training labels}
\KwTarget{optimize the parameters of query encoder $\varphi$.}
Fetch a batch of training queries $\{q_i\}$. \\
Encode queries to embeddings $\{\varphi(q_i)\}$. \\
Use the query embeddings $\{\varphi(q_i)\}$ and the index $I_{doc}$ to retrieve the top-n documents, $\{D_i\}=\{[doc_{i,j}]\}$ ($1$$\leq$$j$$\leq$$n$, $j$ is the ranking position).  \\
If $D_i$ are all labeled irrelevant, replace the last document in the list with a relevant document. \\
Lookup $\{D_i\}$ representations from $E_{doc}$. \\
Compute the relevance scores with inner product. Assume $r_{i,j}$ is $q_i$ and $doc_{i,j}$'s relevance score. \\
Compute loss $\{[loss_i(s,t)]\}$ ($1$$\leq$$s,t$$\leq$$n$) with pairwise loss function $\mathcal{L}$. Assume $l_{i,j}$ is $q_i$ and $doc_{i,j}$'s relevance label. Thus $loss_i(s,t)$ is formulated as follows:
\[
loss_i(s,t) = \left\{
    \begin{array}{ll}
        \mathcal{L}(r_{i,s}, r_{i,t}, s, t) & l_{i,s} > l_{i,t} \\
        0 & l_{i,s} \leq l_{i,t}
    \end{array}
\right. % = 
\]\\
Back propagate the gradients and update the $\varphi$ parameters. \\
Repeat Steps 1 to 8 till convergence.
\caption{Learning To Retrieve}
\label{algo:LTRe}
\end{algorithm}

According to the algorithm design, LTRe has the following advantages. 
First, it is consistent in training and inference because it optimizes the model based on full retrieval results. 
Second, it utilizes the top-ranked results as hard negatives. This would help the model not only filter out the easy negatives but also select the true relevant results from the hard negatives that are somewhat relevant or related to the query.
Third, it computes and deploys the document index once for all, which brings additional efficiency benefits compared with the iterative index refreshes~\cite{guu2020realm, xiong2020approximate}.
Forth, it can use a compressed document index to further reduce the computational costs.

\subsection{Loss Functions}
We adopt two pairwise loss functions, RankNet and LambdaRank~\cite{burges2010ranknet}. In our experiments, we find that their performance varies in different situations. Thus, we select the better one according to the performance on the development set.

Before further elaboration, we restate several denotations. Given a query $q_i$, we retrieve documents from the entire corpus. The predicted relevance score for $j_{th}$ document is $r_{i,j}$. The pairwise loss functions take two documents as input. We assume the two documents are ranked at the $s_{th}$ and $t_{th}$ positions, and the $s_{th}$ document is labeled as more relevant.

\subsubsection{RankNet}

RankNet is a simple pairwise loss that only considers the relevant scores and ignores the ranking positions. It trains the DR model to predict higher scores for more relevant documents, which is formulated as follows: 

\begin{equation}
    \mathcal{L}_{RankNet}(r_{i,s}, r_{i,t}, s, t) = \mathcal{L}_{RankNet}(r_{i,s}, r_{i,t}) = {\rm log}(1+e^{r_{i,t}-r_{i,s}})
\end{equation}

\subsubsection{LambdaRank}
Compared with RankNet, LambdaRank additionally considers the ranking positions and the optimization of evaluation metrics. Given a metric $M$ and two documents ranked at the $s_{th}$ and $t_{th}$ positions, it swaps the two documents in the ranking list and measures the absolute performance change, $| \Delta M(s,t) |$. It then multiplies the change with RankNet loss value as follows:

\begin{equation}
    \mathcal{L}_{LambdaRank}(r_{i,s}, r_{i,t}, s, t) = | \Delta M (s,t) | \times \mathcal{L}_{RankNet}(r_{i,s}, r_{i,t}) 
\end{equation}

\section{Experimental Settings}
\label{sec:exp_setting}

This section describes our experimental setups. Note that for simplicity of expression, in the following sections, the DR training method abbreviation may also refer to a DR model trained with this method. For example, LTRe may also refer to a DR model trained with LTRe. The specific meaning depends on the context. 

\subsection{Datasets}

We conduct experiments on the TREC 2019 Deep Learning (DL) Track~\cite{craswell2020overview}. The Track focuses on ad-hoc retrieval and consists of the passage retrieval and document retrieval tasks. Each task has a fixed document corpus. The retrieval system aims to satisfy the user's information need by returning relevant documents based on the input queries. The queries are sampled from Bing's search logs. NIST accessors label the test sets on the top-ranked results from Track participants. The detailed statistics are shown below:

\begin{itemize}
	\item Passage Retrieval: The passage retrieval task has a corpus of 8.8 million passages with 503 thousand training queries, 7 thousand development queries, and 43 test queries. 
	\item Document Retrieval: The document retrieval task has a corpus of 3.2 million documents with 367 thousand training queries, 5 thousand development queries, and 43 test queries. 
\end{itemize} 

We follow the official settings of TREC 2019 DL Track. Thus the results are comparable with all TREC participating runs. We present MRR@10 and Recall@1k on MSMARCO Passage Dev and NDCG@10 on TREC passage and document test sets.

\subsection{Baselines}

\subsubsection{Sparse Retrieval} 
We list several representative results according to TREC overview paper~\cite{craswell2020overview} and runs on MS MARCO~\cite{bajaj2016ms} leaderboard, such as standard BM25, the best traditional retrieval method, BERT weighted BM25 (DeepCT)~\cite{dai2019context}. We also list cascade systems with BM25 as first-stage retriever, such as the best LeToR, and BERT Reranker.

\subsubsection{Dense Retrieval} 
\label{sec:DR_baseline} 
All DR models in this paper use the same architecture introduced in Section~\ref{sec:overview}. Following pervious works~\cite{luan2020sparsedense, xiong2020approximate}, We initialize the query encoder and document encoder with $\text{RoBERTa}_\text{base}$~\cite{liu2019roberta} and add a $768\times768$ projection layer on top of the last layer's ``[CLS]'' token, followed by a layer norm.

We present DR baselines trained with different negative sampling methods, including random samples from the entire corpus (Rand Neg)~\cite{huang2020embedding}, BM25 top candidates (BM25 Neg)~\cite{gao2020complementing}, Noise Contrastive Estimation, which is the highest scored negatives in batch (NCE Neg)~\cite{gutmann2010noise}, and the 1:1 combination of BM25 and Random negatives (BM25 + Rand Neg)~\cite{luan2020sparsedense, karpukhin2020dense}. We also provide baselines initialized with trained BM25 Neg model and then further trained with other negative sampling methods. They are denoted as BM25 $\rightarrow *$. The DR baselines also include ANCE~\cite{xiong2020approximate}, the previous state-of-the-art DR model. It also uses the BM25 Neg model for initialization. For most baselines, we show the performance reported by Xiong et al.~\cite{xiong2020approximate}. We re-evaluate BM25 Neg and ANCE based on their open-source models. The results are marginally different maybe due to possible environmental differences.

\subsubsection{End-to-End Retrieval}
Recently, Khattab et al.~\cite{Khattab2020ColBERTEA} proposed a competitive end-to-end retrieval method, ColBERT. Its retrieval performance almost matches existing BERT-based two-stage models and outperforms many non-BERT baselines. We list the retrieval performance and latency reported in the original paper.

\subsection{LTRe}
We present LTRe(BM25 Neg) and LTRe(ANCE), which use BM25 Neg model and ANCE~\cite{xiong2020approximate} as the pretrained document encoders, respectively. The pretrained document encoder is also used to initialize the parameters of the query encoder. 
We use a widely-adopted similarity search library, Faiss~\cite{johnson2019billion}, to build the DR index. The default DR index in this paper is IndexFlatIP, an uncompressed DR index that performs inner product search. We also conduct experiments with different compressed DR indexes, denoted as OPQ$n$,IVF1,PQ$n$ where smaller $n$ indicates more compression. Following Dai et al.~\cite{dai2019context}, we use only the first 512 tokens for long documents. 

LTRe uses the same hyperparameters in both passage retrieval and document retrieval tasks. We utilize AdamW~\cite{loshchilov2017decoupled} with the initial learning rate set to $5 \times 10^{-6}$, $\beta_1 = 0.9$, $\beta_2 = 0.999$, L2 weight decay of $0.01$, batch size of $32$, learning rate warm up over the first $2,000$ steps, and linear decay of the learning rate. We use a dropout probability of $0.1$ in all layers.

\section{Results And Discussion}
\label{sec:results_discuss}
This section discusses the experimental results. We first investigate the training process in Section~\ref{sec:exp_training_investigate}. Then we show the effectiveness of LTRe in Sections~\ref{sec:exp_first_stage_effective} and \ref{sec:exp_comparison_with_e2e}. We discuss the training efficiency in Section~\ref{sec:exp_train_efficiency}. 
\subsection{Training Process}
\label{sec:exp_training_investigate}

\begin{figure*}
    % Maximum length
    \hspace*{\fill}%
    \subcaptionbox{The MRR@10 at each training step.
    \label{fig:training_mrr10_curve}}
    {\includegraphics[width=.48\linewidth]{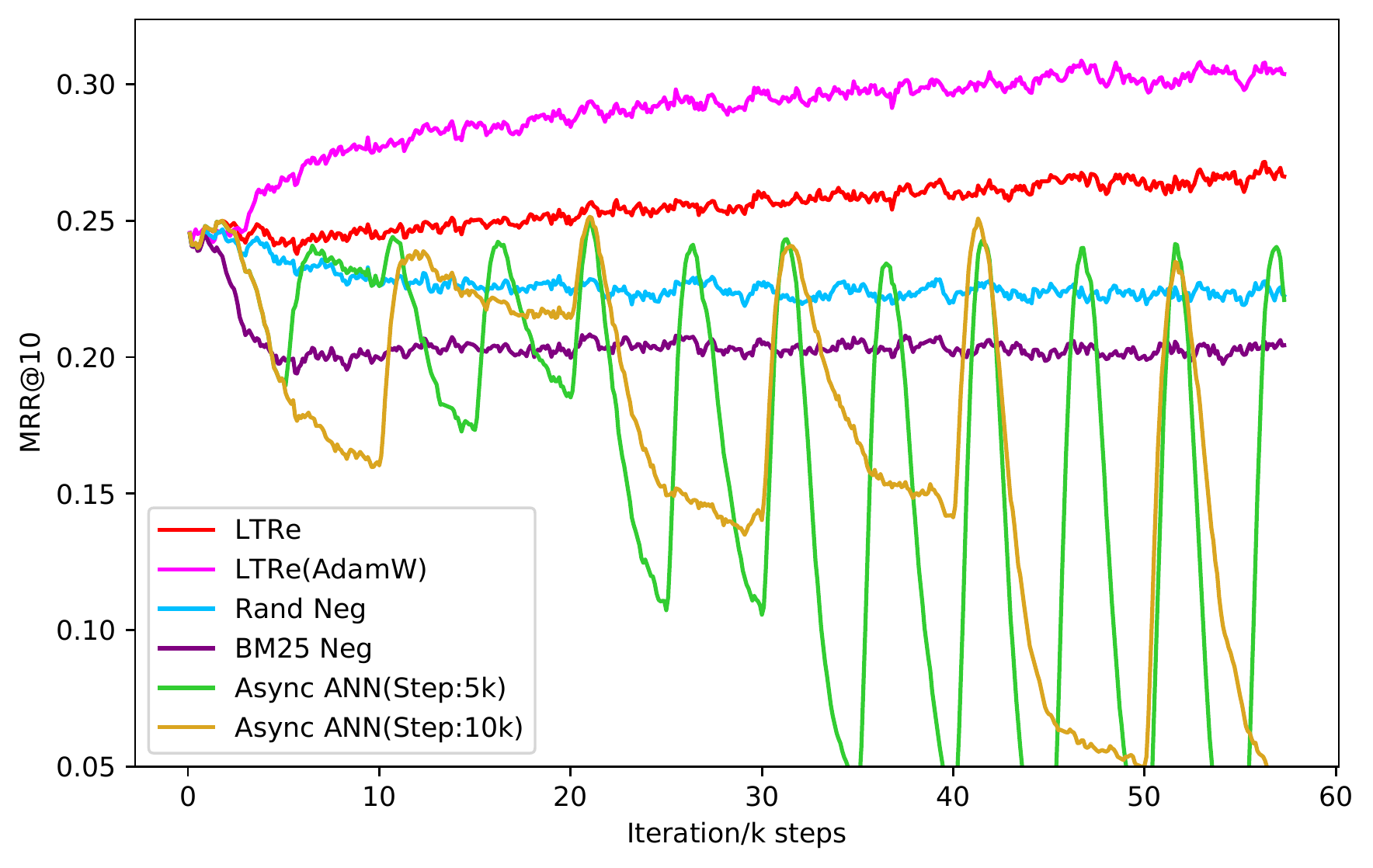}}
    \hfill%
    \subcaptionbox{The Recall@200 at each training step.
    \label{fig:training_random_recall}}
    {\includegraphics[width=.48\linewidth]{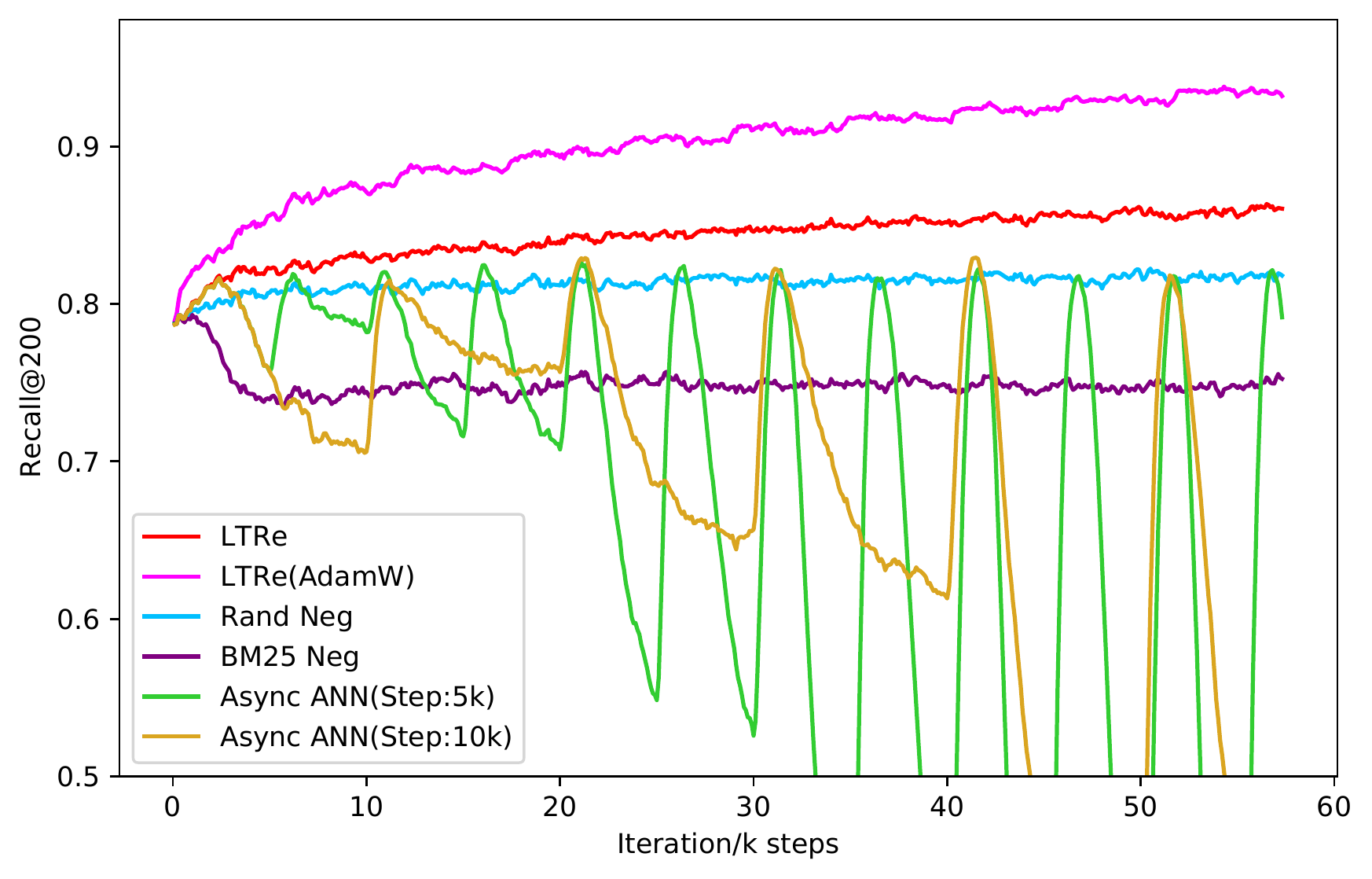}}%
    \hspace*{\fill}%
    \bigskip
  
    % Equal length
    \hspace*{\fill}%
    \subcaptionbox{The overlap of BM25 top-200 and the dense retrieved top-200. For example, the overlap is 0.1 if there are 20 documents retrieved by both BM25 and DR.
    \label{fig:training_bm25_overlap}}
    {\includegraphics[width=.48\linewidth]{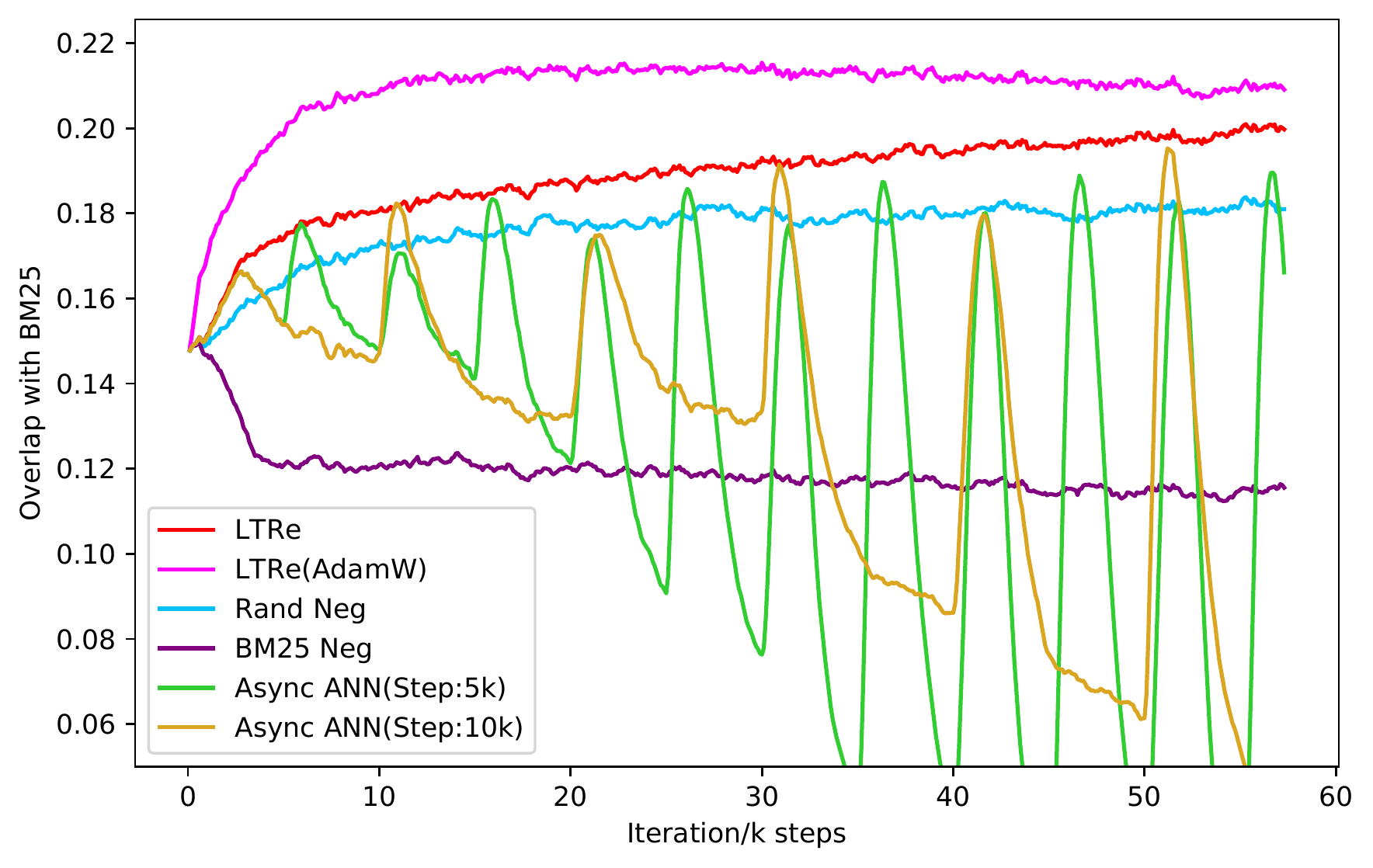}}%
    \hfill%
    \subcaptionbox{The overlap of asynchronous retrieved top-200 and the real-time retrieved top-200. Training methods except for Async ANN use the asynchronous retrieval frequency of 10k steps.
    \label{fig:training_ance_overlap}}
    {\includegraphics[width=.48\linewidth]{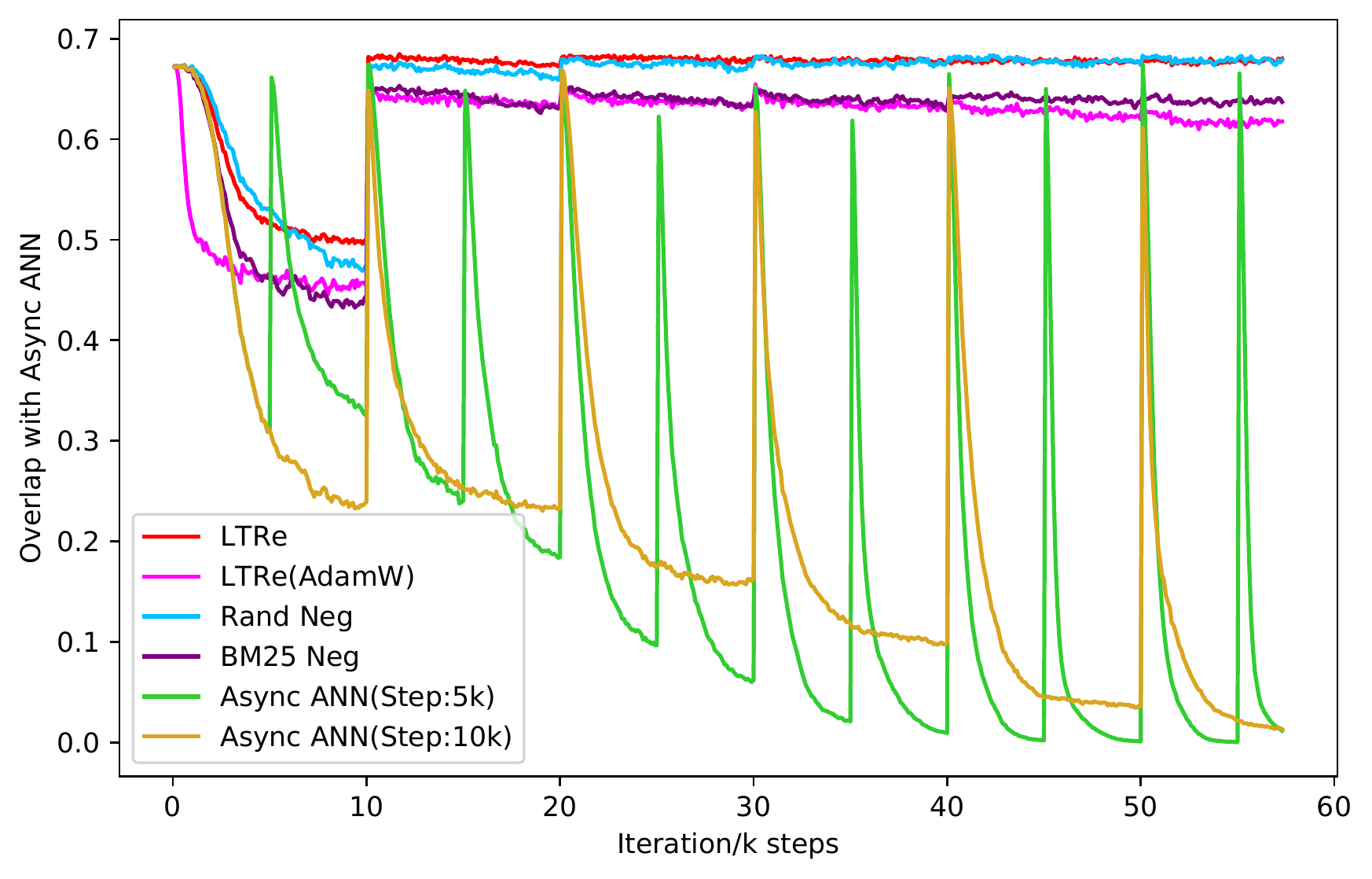}}
    \hspace*{\fill}%
    \caption{The investigation of training procedure on TREC 2019 DL Track document retrieval task. At each training step, the DR model represents a batch of training queries with embeddings. We perform full retrieval with the embeddings and compute different metrics based on the retrieval results. X-axes is the training steps in thousands.}
\end{figure*}

We first investigate whether LTRe can effectively improve the retrieval performance through training. We evaluate the retrieval performance with two evaluation metrics, MRR@10 and Recall@200. MRR@10 focuses on the performance of retrieving top documents, and Recall@200 measures whether the retriever can recall most relevant documents. 
We plot the changes of these two evaluation metrics during the training process. 

For negative sampling baselines, we use BM25 top documents (BM25 Neg), random samples from the entire corpus (Rand Neg), and asynchronous approximate nearest neighbors (Async ANN) with two refresh frequencies. For LTRe, we report the results of a model using the same Lamb~\cite{you2019large} optimizer as other baselines. We also report the results using the AdamW~\cite{loshchilov2017decoupled} optimizer, which is more suitable for LTRe but not for the baselines.

We conduct the experiment on TREC 2019 DL Track document retrieval task. All models use the BM25 Neg model trained on passage retrieval task as a warmup checkpoint. We pre-build the document index and only train the query encoder. Thus, we can evaluate the model with the current batch of training queries at each training step. Specifically, we perform full retrieval using the computed representations of the current batch of queries and calculate the evaluation metrics with the retrieved results. Other training settings are the same as those introduced in Section~\ref{sec:DR_baseline}. 

Figure~\ref{fig:training_mrr10_curve} shows the model's top retrieval accuracy (MRR@10) at each training step. From the results, we find that LTRe brings a fast and steady improvement in MRR@10. Other negative sampling methods, on the contrary, fail to train the model effectively. Rand Neg and BM25 Neg even make the performance worse than the warmup checkpoint. Async ANN is very unstable, and the performance changes dramatically and periodically. 
 
Figure~\ref{fig:training_random_recall} shows the trends of Recall@200 metric. Similar to the results shown in Figure~\ref{fig:training_mrr10_curve}, LTRe provides rapid and steady optimization; BM25 Neg fails to optimize the evaluation metric; Async ANN's performance is very unstable. The only difference is that Rand Neg slightly improves the recall. It is consistent with Huang et al.~\cite{huang2020embedding} in which random negative samples are used for the recall optimization task.

In the following, we analyze different training methods separately according to the experimental results.

\textbf{LTRe:} 
Figures~\ref{fig:training_mrr10_curve} and \ref{fig:training_random_recall} show that LTRe steadily improves the DR model's retrieval performance during training. It outperforms all baselines in both MRR@10 and Recall@200. LTRe effectively teaches the DR model how to retrieve.

\textbf{Rand Neg:} 
From Figures~\ref{fig:training_mrr10_curve} and \ref{fig:training_random_recall}, we can see that Rand Neg learns to roughly judge relevance but cannot effectively select the most relevant documents from the corpus. 
Since most documents in the corpus are not relevant at all to a given query, the random samples are very weak negatives. The DR model learns to distinguish the relevant documents from these easy negatives. Thus, Rand Neg can improve Recall@N when N is large (like 200). 
However, it does not learn to select the most relevant document from a group of relevant candidates. Therefore, it fails to optimize the precision-oriented metric like MRR@10. 

\textbf{BM25 Neg:} Further investigation indicates the optimization of BM25 Neg is biased, which we believe causes its bad performance in Figures~\ref{fig:training_mrr10_curve} and \ref{fig:training_random_recall}. We investigate DR model's consistency with BM25, which is measured with their overlap of top-200 documents. We show the results in Figure~\ref{fig:training_bm25_overlap}.
According to the results, LTRe and Rand Neg improve the overlap with BM25. It agrees with Dai et al.~\cite{Dai2020ContextAwareDT} that suggest the BM25 score is instructive for training neural retrievers.
However, BM25 Neg significantly reduces consistency. The reason is as follows. BM25 top documents are utilized as negatives during training. Thus the DR model learns not to retrieve documents with much query term overlapping, which is a distinct characteristic of these negative examples. Such behavior leads to optimization bias and harms the retrieval performance.

\textbf{Async ANN:} Async ANN has similar optimization bias and thus yields unstable optimization. 
As introduced in Section~\ref{sec:existing_training_methods}, Async ANN retrieves documents for training queries every five thousand or ten thousand training steps. The retrieved top documents serve as negatives for the following training iterations until the next refresh. Though the negatives are based on out-dated model parameters, Async ANN assumes that with high refresh frequency~\footnote{
ANCE~\cite{xiong2020approximate} found that refreshing every 5k steps yields better convergence but requires so many GPUs that they finally chose 10k as refresh frequency.}
, the pre-retrieved negatives can approximate the real-time retrieved top documents. 
To verify this assumption, we compute their overlap during training. As a comparison, We also measure the overlap for other baselines and LTRe, though they do not use the asynchronous retrieved documents as negatives. The results are shown in Figure~\ref{fig:training_ance_overlap}~\footnote{
The dropout operation adds random noise during training. Even based on the same model parameters, the output query representations are different in training and evaluation. Thus, the overlap is less than $100\%$ even for the training step immediately after the Async ANN refresh.}. 
The overlap is high if training methods do not use the pre-retrieved documents as negatives, such as LTRe, Rand Neg, and BM25 Neg. However, for Async ANN, the overlap jumps immediately after each retrieval and then quickly drops. Thus the results do not support Async ANN's assumption. Like BM25 Neg, Async ANN teaches the DR model not to retrieve the documents it retrieved a few thousand steps before. Thus, the optimization is biased and unstable~\footnote{
ANCE~\cite{xiong2020approximate} utilized Async ANN but also tied the parameters of the query encoder and document encoder together. The tie may provide a regularization term and help stabilize the training, but it cannot fundamentally resolve the problem of Async ANN.}.

\subsection{First-Stage Effectiveness}
\label{sec:exp_first_stage_effective}

\begin{table*}[t]
    \centering
    \small
    \begin{tabular}{l|llll|ll} \hline \hline
    & \multicolumn{2}{c}{\textbf{MARCO Dev}} 
    & \multicolumn{2}{c|}{\textbf{TREC DL Passage}} 
    & \multicolumn{2}{c}{\textbf{TREC DL Document}}
    \\ 
    
      & \multicolumn{2}{c}{\textbf{Passage Retrieval}} 
      & \multicolumn{2}{c|}{\textbf{NDCG@10}} 
      & \multicolumn{2}{c}{\textbf{NDCG@10}} 
    \\  \hline
    &  \textbf{MRR@10} &  \textbf{Recall@1k}
      & \textbf{Rerank} & \textbf{Retrieval}
     & \textbf{Rerank} & \textbf{Retrieval}
    \\ \hline
    
\textbf{Sparse \& Cascade IR}  &  &  &  &  & &  \\
{BM25} & 0.187 & 0.858 & -- & 0.506 & -- & 0.519 \\
{Best TREC Trad Retrieval} & n.a. & n.a. & -- & 0.554 & -- & 0.549\\
{DeepCT}~\cite{dai2019context} & 0.243 & 0.910 & -- & n.a. & -- & 0.554\\
{Best TREC Trad LeToR} & n.a. & -- & 0.556 & --   & 0.561 & -- \\ 
{BERT Reranker}~\cite{nogueira2019passage} & 0.365  & -- & \textbf{0.742} & -- & 0.646 & -- 
\\\hline

{\textbf{Dense Retrieval}}  &  &  &  &  & & \\
Rand Neg & 0.261 & 0.949 & 0.605 & 0.552 & 0.615 & 0.543 \\
NCE Neg~\cite{gutmann2010noise} & 0.256 & 0.943 & 0.602 & 0.539 & 0.618 & 0.542 \\
BM25 Neg~\cite{gao2020complementing} & 0.309 & 0.935 & 0.677 & 0.607 & 0.641 & 0.539 \\ 
BM25 + Rand Neg~\cite{karpukhin2020dense, luan2020sparsedense} & 0.311 & 0.952 & 0.653 & 0.600 & 0.629 & 0.557\\
BM25 $\rightarrow$ Rand & 0.280 & 0.948 & 0.609 & 0.576 & 0.637 & 0.566 \\
BM25 $\rightarrow$ NCE Neg & 0.279 & 0.942 & 0.608 & 0.571 & 0.638 & 0.564 \\
BM25 $\rightarrow$ BM25 + Rand & 0.306 & 0.939 & 0.648 & 0.591 & 0.626 & 0.540 \\
{ANCE (Xiong et al.)}~\cite{xiong2020approximate} & 0.330 & 0.959 & 0.677 & 0.648 & 0.641 & 0.615 \\ 
{ANCE (Ours)}~\cite{xiong2020approximate} & 0.338 & 0.960 & 0.681 & 0.654 & 0.644 & 0.610 \\ 

\hline 
{LTRe (BM25 Neg)} & 0.329$\ssymbol{1}$ & 0.955$\ssymbol{1}$ & 0.685 & 0.661$\ssymbol{1}$  & 0.649 & 0.610$\ssymbol{1}$ \\ 
{LTRe (ANCE)} & \textbf{0.341}$\ssymbol{1}{}\ssymbol{2}$  & \textbf{0.962}$\ssymbol{1}{}\ssymbol{2}$ & 0.687 & \textbf{0.675}$\ssymbol{1}{}\ssymbol{2}$ & \textbf{0.663}$\ssymbol{1}{}\ssymbol{2}$ & \textbf{0.634}$\ssymbol{1}{}\ssymbol{2}$ \\ 
\hline 
    \end{tabular}
     \caption{Results on TREC 2019 Deep Learning Track. $\ssymbol{1}$ and $\ssymbol{2}$ indicate LTRe's statistically significant improvements over BM25 Neg and ANCE, respectively. We use paired t-test with p-value threshold of 0.05. Best results in each category are marked bold. Results not available and not applicable are marked as ``n.a.'' and ``--'', respectively. 
\smallskip~\label{tab:first_stage_effectiveness}}
\end{table*}

This section presents the first-stage effectiveness of LTRe based on TREC 2019 DL track passage retrieval task and document retrieval task. The results are shown in Table~\ref{tab:first_stage_effectiveness}. 

\subsubsection{Sparse Retrieval Baselines} 
Traditional sparse retrievers are very strong on the passage and document test sets. The Best TREC Traditional retriever matches several DR models, such as Rand Neg, NCE Neg, BM25 Neg. DeepCT~\cite{dai2019context} uses BERT to weight terms and further improves bag-of-words models. Traditional LeToR reranker provides slight performance improvement, while BERT reranker significantly improves the performance.

\subsubsection{Dense Retrieval Baselines}  
Simple negative sampling methods, such as Rand Neg, NCE Neg, and BM25 Neg, cannot yield consistent or significant improvement compared with traditional retrievers. In general, they perform better on passage retrieval than document retrieval. 

Using BM25 Neg as warmup brings additional improvement. For example, BM25 $\rightarrow$ Rand significantly improves Rand Neg's top retrieval accuracy. 
Particularly, initialized with BM25 Neg, ANCE achieved previous state-of-the-art results. It outperforms all other sparse and dense retrieval baselines.

Several results validate our training process investigation. Section~\ref{sec:exp_training_investigate} concludes that Rand Neg teaches the DR model to judge relevance roughly. In Table~\ref{tab:first_stage_effectiveness}, Rand Neg performs the worst in top retrieval accuracy, but it significantly improves Recall@1k compared with other DR baselines, such as BM25 Neg, NCE Neg, and BM25 $\rightarrow$ NCE. Section~\ref{sec:exp_training_investigate} concludes that BM25 Neg is biased and mainly teaches the DR model how to rerank BM25 top candidates. In Table~\ref{tab:first_stage_effectiveness}, BM25 Neg significantly improves reranking performance but performs the worst in Recall@1k. For example, on MSMARCO Dev passage retrieval, BM25 Neg significantly outperforms Rand Neg in MRR@10 (0.309 vs. 0.261), but underperforms it in Recall@1k (0.935 vs. 0.949).

\subsubsection{LTRe} 
LTRe effectively improves the full retrieval performance for both passage retrieval task and document retrieval task. 

LTRe(BM25 Neg) achieves similar performance with ANCE and outperforms all other sparse and dense retrieval baselines. 
Considering that LTRe does not optimize the document encoder like all DR baselines, such results are promising. 
The DR baselines aim to optimize both document representations and query representations. They teach the DR model how to rerank document samples.
On the contrary, LTRe fixes the document representations so it can perform full retrieval during training. It optimizes the query encoder and forces the model better represents the user's information need. 
The optimization is based on the full retrieval results and the model directly learns to retrieve rather than to rerank.
The experimental results show that LTRe successfully improves the first-stage retrieval performance.

We plot a t-SNE example in Figure~\ref{fig:tsne} using LTRe(BM25 Neg) model and a query from the TREC 2019 DL Track document retrieval test set. BM25 Neg model generates the document representations and the initial query representation. It fails to accurately retrieve relevant documents for this query.
After training, LTRe(BM25 Neg) model understands the user's information need. It maps the query closer to the relevant documents. Thus, they will be retrieved in higher positions and the retrieval performance is improved.

LTRe(ANCE) achieves the best results and significantly outperforms the previous state-of-the-art ANCE model. It suggests that a better document encoder yields better retrieval performance. 
ANCE utilizes Async ANN as negatives, which results in biased and unstable training, as discussed in Section~\ref{sec:exp_training_investigate}. The training may be sensitive to hyperparameters, and the researchers may need to carefully select checkpoints after training. 
Our experimental results show that while ANCE can optimize the document encoder, it fails to find the optimal parameters of the query encoder. Our proposed method, LTRe, further improves its performance by a large margin through fine-tuning the query encoder.

The performance of LTRe(ANCE) nearly matches BM25-BERT two-stage retrieval system on the document retrieval task. Section~\ref{sec:exp_comparison_with_e2e} shows that on the passage retrieval task, LTRe(ANCE) significantly outperforms BM25-BERT under reasonable latency constraints. 
Thus, the representation-based~\cite{guo2019deep} model's retrieval performance can match the interaction-based cascade retrieval system or even outperform it under some latency constraints.
Such results may prompt researchers to reconsider the necessity of modeling interactions between query and document terms. 
It provides additional performance gains but also a substantial time overhead. 
Representation-based models, conversely, are very promising considering their efficiency advantage and good retrieval performance.

\subsection{Two-Stage Effectiveness}
\label{sec:exp_comparison_with_e2e}

\begin{table}[t]
    \centering
    \small
    \begin{tabular}{l|ccc} \hline \hline
    & \textbf{ReRank Depth}
    & \textbf{MRR@10}
    & \textbf{Latency (ms)}
    \\ \hline
    
{\textbf{End-to-End}} & & & \\
{ColBERT}~\cite{Khattab2020ColBERTEA} & - & 0.360 & 458  \\ 

\hline
{\textbf{First-Stage}} & & &  \\
{BM25}~\cite{yang2018anserini}  & - & 0.190 & 62 \\
{LTRe}  & - & 0.341 & 47 \\

\hline
{\textbf{BM25 Two-Stage}} &     &  &  \\
{BM25 + $\text{BERT}_\text{base}$} &  10 & 0.275 & 95 \\
{BM25 + $\text{BERT}_\text{base}$} &  30 & 0.320 & 165 \\
{BM25 + $\text{BERT}_\text{large}$} & 10 & 0.279 & 164 \\
{BM25 + $\text{BERT}_\text{large}$} & 30 & 0.323 & 393 \\
\hline
{\textbf{LTRe Two-Stage}}  &    &  &  \\
{LTRe + $\text{BERT}_\text{base}$}  & 10 & 0.358 & 79 \\
{LTRe + $\text{BERT}_\text{base}$}  & 30 & 0.367 & 149 \\
{LTRe + $\text{BERT}_\text{large}$} & 10 & 0.362 & 148 \\
{LTRe + $\text{BERT}_\text{large}$} & 30 & \textbf{0.375} & 375 \\

\hline 
    \end{tabular}
     \caption{Comparison between End-to-End retrieval and Two-Stage retrieval on MS MARCO Dev passage retrieval dataset. LTRe is short for LTRe(ANCE).
     \smallskip~\label{tab:comparison_with_e2e}}
\end{table}

This section presents the two-stage effectiveness of LTRe. We compare LTRe(ANCE)+BERT, BM25+BERT, and a competitive end-to-end retrieval baseline, ColBERT~\cite{Khattab2020ColBERTEA}. The performance and latency are shown in Table~\ref{tab:comparison_with_e2e}. This section uses LTRe as the abbreviation for LTRe(ANCE). 

The evaluation details are as follows. Khattab et al.~\cite{Khattab2020ColBERTEA} evaluated ColBERT on an advanced 32GB Tesla V100 GPU. We present the latency and retrieval performance they reported. We evaluate LTRe on three 11GB GeForce RTX 2080 Ti GPUs so the entire index can be loaded into GPU memory. We evaluate BERT on a single 11GB GeForce RTX 2080 Ti GPU. We finetune our $\text{BERT}_\text{base}$ model following Nogueira et al.~\cite{nogueira2019passage} and adopt their open-source $\text{BERT}_\text{large}$ model.

According to Table~\ref{tab:comparison_with_e2e}, LTRe's first-stage retrieval performance outperforms BM25+BERT under reasonable efficiency constraints. With BERT as the second-stage ranker, LTRe outperforms ColBERT in both effectiveness and efficiency. 
Precisely, reranking LTRe top 10 documents with $\text{BERT}_\text{base}$ almost matches ColBERT and is about six times faster. Reranking more candidates or reranking with $\text{BERT}_\text{large}$ outperforms ColBERT and is three times faster. The combination, reranking 30 candidates with $\text{BERT}_\text{large}$, provides $4\%$ performance improvement against ColBERT and is still faster.

The results show that LTRe has the following advantages. 
First, compared with the end-to-end retrieval system, LTRe retrieves documents much faster. Thus it can benefit from a powerful second-stage ranker to further improve performance. Experiments show that LTRe+BERT outperforms ColBERT in both effectiveness and efficiency.
Second, compared with traditional retrievers, LTRe significantly improves the reranking performance. Experiments show that LTRe+BERT greatly outperforms BM25+BERT.
Third, LTRe can even be directly used without a second-stage ranker. Experiments show that LTRe's retrieval performance significantly outperforms BM25+BERT under reasonable efficiency constraints.

\subsection{Training Efficiency}
\label{sec:exp_train_efficiency}

\begin{table*}[t]
    \centering
    \small
    \begin{tabular}{l|cc|ccc|ccc} \hline \hline
    & \multicolumn{2}{c|}{\textbf{Index}}
    & \textbf{GPU}
    & \multicolumn{2}{c|}{\textbf{Training}}
    & \multicolumn{2}{c}{\textbf{MARCO Dev}} 
    & \textbf{TREC DL}
    \\ 
      & \textbf{Memory} & \textbf{Quality}
      & \textbf{Resources}
	  & \multicolumn{2}{c|}{\textbf{time}}
      & \multicolumn{2}{c}{\textbf{Passage Retrieval}} 
      & \textbf{NDCG@10} 
    \\  \hline
    & \textbf{GB} & \textbf{MRR@10} 
    & \textbf{2080 Ti}
    & \textbf{Hours} & \textbf{Speedup} 
    &  \textbf{MRR@10} &  \textbf{Recall@1k}
    & \textbf{Retrieval}
    \\ \hline
    
{\textbf{Baselines}} & & & &  &  &  &  & \\
{BM25 Neg}~\cite{gao2020complementing} & 4.2 & 0.167 & - & - & - & 0.309 & 0.935 & 0.607 \\ 
{ANCE (Ours)}~\cite{xiong2020approximate} & 26 & 0.309 & 4 & 645 & 1x & \textbf{0.338} & 0.960  & 0.654 \\ 
\hline
{\textbf{LTRe (BM25 Neg)}}  &  &  &  &  & \\
{OPQ6,IVF1,PQ6}   & 0.1 & 0.050 & - & -   & -    & 0.304 & 0.946  & 0.627 \\  
{OPQ12,IVF1,PQ12} & 0.2 & 0.151 & - & -   & -    & 0.318 & 0.948  & 0.635 \\  
{OPQ24,IVF1,PQ24} & 0.2 & 0.221 & 1 & 3.0 & 215x & 0.324 & 0.949  & 0.644 \\  
{OPQ48,IVF1,PQ48} & 0.5 & 0.254 & 1 & 3.2 & 202x & 0.327 & 0.950  & 0.652 \\  
{OPQ96,IVF1,PQ96} & 0.9 & 0.273 & 1 & 3.7 & 174x & 0.326 & 0.953 & 0.656 \\  
{IndexFlatIP}       & 26  & 0.309 & 4 & 3.6 & 179x & 0.329 & \textbf{0.962} & \textbf{0.661} \\  

\hline 
    \end{tabular}
     \caption{Training efficiency comparison between LTRe and ANCE on TREC 2019 DL passage retrieval task. We use different indexes for LTRe training. IndexFlatIP is an uncompressed DR index. OPQ$n$,IVF1,PQ$n$ is a compressed DR index with hyperparameter $n$. Smaller $n$ corresponds to more compression. The compressed index with $n=6$ or $n=12$ is not supported on GPU. Thus we do not compare their training speed with other GPU accelerated methods.
\smallskip~\label{tab:training_efficiency}}
\end{table*}

\begin{table*}[t]
    \centering
    \small
    \begin{tabular}{l|cccc} \hline \hline
    \textbf{NDCG@10}
    & \textbf{OPQ24,IVF1,PQ24} 
    & \textbf{OPQ48,IVF1,PQ48} 
    & \textbf{OPQ96,IVF1,PQ96} 
    & \textbf{IndexFlatIP} 

    \\ \hline
 
{\textbf{LTRe (BM25 Neg)}} & & & & \\
{OPQ24,IVF1,PQ24} & \textbf{0.552} & 0.587 & 0.615 & 0.644 \\  
{OPQ48,IVF1,PQ48} & 0.542 & \textbf{0.597} & 0.619 & 0.652 \\  
{OPQ96,IVF1,PQ96} & 0.536 & 0.581 & \textbf{0.624} & 0.656 \\  
{IndexFlatIP}       & 0.536 & 0.586 & 0.622 & \textbf{0.661} \\  

\hline 
    \end{tabular}
     \caption{NDCG@10 results on test set of TREC 2019 DL Passage retrieval task. Each row and each column correspond to using different compressed indexes to train and evaluate, respectively. IndexFlatIP is an uncompressed DR index. OPQ$n$,IVF1,PQ$n$ is a compressed DR index with hyperparameter $n$. Smaller $n$ corresponds to more compression. 
\smallskip~\label{tab:eval_different_indexes}}
\end{table*}

As introduced in Section~\ref{sec:principles}, an ideal training mechanism should be applicable to a large corpus. Since ANCE~\cite{xiong2020approximate} achieved the previous state-of-the-art results, this section compares its training efficiency with our proposed LTRe. We discuss from two aspects, namely training time and computing resources.

The evaluation details are as follows. We measure the training time with 11GB GeForce RTX 2080 Ti GPUs. 
We use Product Quantization~\cite{jegou2010product} to compress the index, which is denoted as OPQ$n$,IVF1,PQ$n$, where smaller $n$ indicates more compression.  
We present the index memory usage and search quality. The search quality is measured with MRR@10 metric on MARCO Dev passage retrieval task. We also evaluate the BM25 index for BM25 Neg model and the uncompressed DR index (IndexFlatIP) for ANCE model. For LTRe and ANCE, the index is evaluated before training, that is, based on the BM25 Neg warmup checkpoint.

The results are shown in Table~\ref{tab:training_efficiency}. 

\subsubsection{Training Time} 
\label{sec:exp_train_efficiency_time}
LTRe provides 179x speed-up in training time against ANCE when using the same uncompressed index, and their performances are similar. LTRe converges in less than 4 hours, but ANCE needs nearly a month. There are two reasons why ANCE is very inefficient.

\begin{itemize}
	\item ANCE iteratively encodes the corpus to embeddings. With three 11GB GeForce RTX 2080 Ti GPUs, encoding once takes 10.75 hours. In contrast, LTRe does not have this overhead because it fixes the document embeddings during training.  
	\item ANCE converges very slowly, which may be caused by the biased and unstable training discussed in Section~\ref{sec:exp_training_investigate}. For example, on the passage retrieval task, ANCE needs 600k steps with batch size of 64 while LTRe needs 60k steps with batch size of 32. 
\end{itemize} 

It is hard to apply ANCE to a large corpus. The time to encode documents increases linearly with the corpus size, which means that the ANCE's inefficiency will significantly worsen if the corpus is larger.  

On the other hand, LTRe's training time is less affected by the corpus size. According to the LTRe training process, the corpus size affects the speed of full retrieval operation. There are two reasons why LTRe is still applicable to a large corpus: First, the full retrieval operation is very efficient and takes a little proportion of the training time. For example, on passage retrieval task with corpus size of 8.8 million, the full retrieval operation takes a total of 40 minutes, which is about $20\%$ of the entire training time. Second, unlike ANCE, the running time of several full retrieval algorithms~\cite{shen2015learning} scale sub-linearly with the corpus size. Thus, a larger corpus does not significantly slow the full retrieval operation down.

\subsubsection{Computing Resources}
\label{sec:exp_train_efficiency_gpu}

This section investigates using a compressed index for training to save computing resources. 

The motivations are as follows.
Dense Retrieval is slow on CPU and relies on GPU to accelerate. To utilize GPU resources, the document index needs to be entirely loaded into GPU memory. However, the uncompressed index is large but the GPU memory is usually limited. Table~\ref{tab:training_efficiency} shows that the uncompressed DR index, IndexFlatIP, is 26GB in size and requires three 11GB GPUs. Thus, it is necessary to study utilizing a compressed index for training.

According to Table~\ref{tab:training_efficiency}, training a DR model with a compressed index greatly saves computing resources and still effectively optimizes the retrieval performance. 
Using a compressed index with $n=96$ reduces the GPU memory footprint to $3\%$ and yields similar retrieval performance with ANCE on TREC 2019 DL test set (0.656 vs. 0.654).
Using a more compressed DR index (smaller $n$) further reduces GPU memory footprint with acceptable performance loss.
For example, a heavily compressed index ($n=12$) reduces the memory footprint to $0.6\%$, and the DR model trained with it still outperforms all sparse and dense retrieval baselines except for ANCE in retrieval performance, according to Tables~\ref{tab:first_stage_effectiveness} and \ref{tab:training_efficiency}.
Note that an over-compressed index ($n=6$) degenerates into random negative sampling. On MARCO Dev, compared with BM25 Neg, it harms the MRR@10 but improves the Recall@1k, which validates our discussion about Rand Neg in Section~\ref{sec:exp_training_investigate}.

We further use different DR indexes to evaluate the DR models trained with different DR indexes, as shown in Table~\ref{tab:eval_different_indexes}. Each row and each column correspond to using a specific DR index to train and evaluate, respectively. 

Table~\ref{tab:eval_different_indexes} shows that the values on the main diagonal are the best performances for this column. Thus, it is best to keep consistency in training and inference. In other words, if a specific index is used in inference, it should also be used in training. Keeping consistency between training and inference is the core idea of LTRe, as shown in Figure~\ref{fig:ltre_flowchart}. Previous negative sampling training methods, as shown in Figure~\ref{fig:negative_flowchart}, ignore this issue.

\section{Conclusion}
\label{sec:conclusion}

In this paper, we present \textbf{L}earning \textbf{T}o \textbf{Re}trieve (\textbf{LTRe}), an effective and efficient training mechanism for Dense Retrieval. 
We verify that previous training strategies train the model to rerank selected document samples rather than to retrieve from the whole indexed corpus. LTRe, however, steadily optimizes full retrieval performance. Experiments show that:
1) In terms of effectiveness, LTRe significantly outperforms all sparse and dense retrieval baselines and even outperforms BM25+BERT cascade system under reasonable efficiency constraints. Compared with a traditional first-stage retriever, it enables the second-stage ranker to yield better performance.
2) In terms of efficiency, LTRe has a better scalability and is applicable for a large-scale corpus. It provides more than 170x speed-up in training time compared with the previous state-of-the-art training method. We can further adopt LTRe with a compressed index, which greatly saves computing resources but only brings a minor loss in retrieval performance.  

There are still some remaining issues for future work. 
First, LTRe achieves good performance by only optimizing the query encoder, but we also find that a better document encoder yields better retrieval performance. Thus, how to pretrain and finetune a document encoder remains to be further explored. 
Second, this paper applies LTRe to ad-hoc retrieval. Future work may examine LTRe in other tasks that require a retrieval module, such as the Open Question Answering task.

\clearpage
\bibliographystyle{unsrt}
\bibliography{references}

\end{document}